\documentclass[prl,amsmath,amssymb,twocolumn, 9pt]{revtex4}

\usepackage{amsthm}
\usepackage{dcolumn}
\usepackage{amsmath}
\usepackage{bm}
\usepackage{graphicx}
\usepackage{xcolor}
\usepackage{ulem}

\input{Qcircuit.tex}
\begin{document}

\newtheorem{corollary}{Corollary}
\newtheorem{definition}{Definition}
\newtheorem{example}{Example}
\newtheorem{lemma}{Lemma}
\newtheorem{proposition}{Proposition}
\newtheorem{theorem}{Theorem}
\newtheorem{fact}{Fact}
\newtheorem{property}{Property}
\newcommand{\braket}[3]{\langle #1|#2|#3\rangle}
%%inner product
\newcommand{\ip}[2]{\langle #1|#2\rangle}
%%outer product
\newcommand{\op}[2]{|#1\rangle \langle #2|}

\newcommand{\tr}{{\rm tr}}
\newcommand {\E } {{\mathcal{E}}}
\newcommand {\F } {{\mathcal{F}}}
\newcommand {\diag } {{\rm diag}}

\title{Five Two-Qubit Gates Are Necessary for Implementing Toffoli Gate}

\author{Nengkun Yu}
\email{nengkunyu@gmail.com}
\author{Runyao Duan}
\author{Mingsheng Ying}

\affiliation{State Key Laboratory of Intelligent Technology and Systems, Tsinghua National Laboratory\protect\\
for Information Science and Technology, Department of Computer Science and Technology,\protect\\
Tsinghua University, Beijing 100084, China}
\affiliation{Center for Quantum Computation and Intelligent Systems (QCIS), Faculty of
Engineering and Information Technology, University of Technology,
Sydney, NSW 2007, Australia}

\begin{abstract}
In this paper, we settle the long-standing open problem of the minimum cost of two-qubit gates for simulating a Toffoli gate. More precisely, we show that five two-qubit gates are necessary. Before our work, it is known that five gates are sufficient and only numerical evidences
have been gathered, indicating that the five-gate implementation is necessary. The idea introduced here can also be used to solve the problem of optimal simulation of three-qubit control phase introduced by Deutsch in 1989.
\end{abstract}

\maketitle

Since quantum computation provides the possibility of solving certain problems which are believed to be infeasible with a classical computer \cite{Sho97,Fey82,Hal02,FKW02}, a huge amount of effort has been devoted to building functional and scalable quantum computers over the last two decades. Quantum logical circuit is the most popular model of quantum computer hardwares. In order to be a general purpose computational device, a quantum computer must implement a small set of quantum logical gates \cite{NC00}, which can universally serve as the basic building blocks of quantum circuits, in the same way as classical logical gates did for conventional digital circuits. It is quite natural to choose certain gates operating on a small number of qubits as the basic gates.

Theoretically, any two-qubit gate that can create entanglement, like the controlled-NOT (CNOT) gate, together with all single-qubit gates is universal \cite{BDDG+2000}. It has also been experimentally demonstrated that two-qubit gates can be realized with high fidelity using the current technology, for example, two-qubit gate with superconducting quibts have been presented with fidelities higher than $90\%$ \cite{DCG+09}. Finding more efficient ways to implement quantum gates may allow small-scale quantum computing
tasks to be demonstrated on a shorter time scale. More precisely, it would be quite helpful for defeating quantum decoherence to realize multi-qubit gates with the least number of possible basic gates. Thus an important problem is how to implement more-than-two-qubit gates using only two-qubit gates. Indeed, studying the minimum cost of two-qubit gates for simulating a multi-qubit gate is not only of theoretical importance, but also an experimental requirement: to accomplish a quantum algorithm, even in a small size, one has to implement a relatively high level of control over the multi-qubit quantum system. A lot of experimental effort has been devoted to demonstrating multi-qubit controlled-NOT gates in ion traps \cite{MKH+09}, linear optics \cite{LBA+09}, superconductors \cite{FSB+12} and atoms \cite{CMP+98}.

The Toffoli gate is perhaps one of the most important quantum logical gates as it can universally realize classical reversible computation \cite{TOF80}, as well as universal quantum computation \cite{SHI03} with little extra help. It also plays a central role in quantum error-correction \cite{CMP+98,KLMN01,CLS+04,PJF05,AOK09}. Recently, experimental implementation of the Toffoli gate has received considerable attention \cite{MKH+09, LBA+09, MWY+11,RDN+12,FSB+12}. However, it remains still unknown what is the optimal simulation of the Toffoli gate by using bipartite quantum logical gates, which has been an important open problem explicitly listed in the influential textbook on quantum computation\cite{NC00}. Here, we settle this problem by showing that five two-qubit gates are necessary and sufficient for implementing the Toffoli gate. A five two-qubit gates decomposition of the Toffoli gate was known long time ago, but before this work numerical evidences showing that the five-gate implementation is optimal have been found \cite{DIV98,DS94,SW95,BBCD+95}. Our result gives, for the first time, a theoretical proof for the optimality beyond numerical evidences.

The function of Toffoli gate is simply a three-qubit controlled NOT gate and can be intuitively explained as follows. The Toffoli gate is acting on three quantum bits, namely  $A$,$B$, and $C$. Here $A$ and $B$ are control qubits, and $C$ is the target qubit. Let us fix a computational basis $\{\ket{0},\ket{1}\}$ for each qubit. Upon an input $\ket{abc}$, the gate will output the states of $A$ and $B$ directly, and flip the system $C$ only if both the states of $A$ and $B$ are $1$. The Toffoli gate can be depicted by the following diagram,
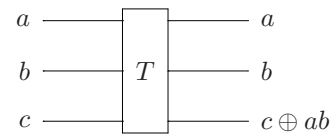
\[
\Qcircuit @C=3em @R=1em {
\lstick{a} & \multigate{2}{T} & \rstick{a}\qw \\
\lstick{b} & \ghost{T}& \rstick{b} \qw\\
\lstick{c} & \ghost{T} & \rstick{c\oplus ab}\qw
}
\]
It can also been written as the following version by considering the output over the computational basis:
\begin{eqnarray*}
T_{ABC}=I-\op{110}{110}-\op{111}{111}+\op{110}{111}+\op{111}{110}.
\end{eqnarray*}
It is well known that the Toffoli gate is universal for the classical computation in the sense that all conventional boolean circuits can be built upon it in a reversible way. It was also proved to be  universal for quantum computation if the one-qubit Hadamard gate is provided as free resource \cite{SHI03}. Furthermore, a series of works showed that the Toffoli gate is an indispensable ingredient in realizing fault$-$tolerant quantum computation \cite{CMP+98,KLMN01,CLS+04,PJF05,AOK09, DEN01}. Recently, a rapid progress has been made on implementing the Toffoli gate experimentally. The first experimental realization of the quantum Toffoli gate is presented in an ion trap quantum computer, in January 2009 at the University of Innsbruck, Austria \cite{MKH+09}. A new approach using higher-dimensional Hilbert spaces was proposed \cite{RRG07} that enables us to simplify the implementation of the Toffoli gate in linear optics \cite{LBA+09} and superconducting circuits \cite{MWY+11,RDN+12,FSB+12}.

Due to its significance in quantum computing, the theoretical pursuit of efficient implementation of the Toffoli gate using a sequence of single- and two-qubit gates has quite long history. It was well known that six CNOT gates are optimal when single-qubit unitary is provided as free resources \cite{MAR94,DIV98,SK04,SM09}. Then an interesting question naturally arises: how many general two-qubit gates rather than the CNOT are required to implement the Toffoli gate? This question has attracted many different researchers in the last two decades. In particular, Nielsen and Chuang explicitly listed it as an unsolved problem in their standard textbook on quantum computation \cite{NC00} (see page 213, Problem 4.4). What we know until now is that the Toffoli gate can be decomposed as a circuit consisting of five two-qubit gates, and numerical evidences have been gathered, indicating that the five-gate implementation is optimal \cite{DIV98,DS94,SW95}. Here, we finally settle this problem and present a theoretical proof of the optimality.

Let $V_{ABC}=I_{ABC}-2\op{111}{111}$ with $I_{ABC}$ the identity operator on Hilbert space $\mathcal{H}_A\otimes \mathcal{H}_B\otimes \mathcal{H}_C$. It is evident that $V_{ABC}=(I_{AB}\otimes H_C) T_{ABC} (I_{AB}\otimes H_C)$, where $H$ is the Hadamard gate given by
\begin{eqnarray*}
H=\frac{1}{\sqrt{2}}\left(
\begin{array}{cc}
 1 & 1 \\
 1 & -1
\end{array}
\right).
\end{eqnarray*}
In other words, $V_{ABC}$ and the Toffoli gate $T_{ABC}$ are equivalent up to local unitary $H_{C}$. By absorbing $H_{C}$ into any two-qubit gates acting on $AC$ or $BC$, we can easily conclude that $V_{ABC}$ and $T_{ABC}$ require the same number of two-qubit gates to realize. Thus in the following discussions, we will focus on the minimal cost of simulating $V_{ABC}$ using two-qubit gates.

The gate $V_{ABC}$ is a real Hermitian matrix that is invariant under any permutation of subsystems $A$,$B$, and $C$. Thus it can be regarded as a controlled-gate with control on each qubit. Note that any bipartite unitary $U_{AB}$ acting on a qubit system $A$ and a general system $B$ is said to be a controlled-gate with control on $A$ if it can be decomposed into the form of $U_{AB}=\op{0_A}{0_A}\otimes U_0+\op{1_A}{1_A}\otimes U_1$.
This simple observation is helpful to reduce the number of cases we need to consider.

Since $V_{ABC}$ is regarded as a three-qubit gate acting on $ABC$, any two-qubit gate used to implement $V_{ABC}$ can be simply classified into three types: $\mathcal{K}_{AB}$ - the gate acting on the systems $A$ and $B$, and likewise, $\mathcal{K}_{BC}$, and $\mathcal{K}_{AC}$. Clearly, it is impossible that all two-qubit gates used to simulate $V_{ABC}$ belong to the same type. Furthermore, we can verify that only two two-qubit gates are not sufficient for the simulation of $V_{ABC}$. To see this, one only needs to notice that $U_{AB}U_{BC}=V_{ABC}$ implies that $U_{BC}$ is also a controlled gate with control system $C$. This leads us to contradiction by a routine calculation.

Three observations are quite helpful during our proof: i). Any two-dimensional two-qubit subspace contains some product state; ii). A two-qubit unitary $U_{AB}$ can be regarded as a controlled gate with control system $A$ if the state of qubit $A$ in $U_{AB}\ket{0}_A\ket{y}_B$ is always $\ket{0}_A$ for any state $\ket{y}_B$ of system $B$; iii) Let $U_{AB}U_{AC}$ be a three-qubit unitary which can be regarded as a controlled gate between the bipartition $A$-$BC$ with control system $A$. Then there exist $v_{B1},v_{B2}$ and $w_{C1},w_{C2}$ being one-qubit gates on $\mathcal{H}_B$ and $\mathcal{H}_C$ such that $U_{AB}U_{AC}=\op{0}{0}\otimes v_{B1}\otimes w_{C1}+\op{1}{1}\otimes v_{B2}\otimes w_{C2}.$

Observations i) and ii) are obvious. To see iii), we can assume $U_{AC}\ket{0}_A\ket{\gamma}_C=\ket{0}_A\ket{\psi}_C$ by moving the local unitary to the left of $U_{AB}$. Then $U_{AB}U_{AC}\ket{0}_A\ket{y}_B\ket{\gamma}_C=U_{AB}\ket{0}_A\ket{y}_B\ket{\psi}_C.$ Note that the state of $A$'s part of $U_{AB}\ket{0}_A\ket{y}_B$ is always $\ket{0}$, which means that $U_{AC}$ is a controlled gate with control on $A$. Similarly, $U_{AB}$ is also a controlled gate with control on $A$. Hence the result follows.

Now we show that three two-qubit gates are not sufficient to implement $V_{ABC}$. We will achieve this goal by analysing all possible circuits consisting of three two-qubit gates. Due to the highly symmetric properties of $V_{ABC}$, we only need to consider the following two cases:\\
Case 1: These three gates belong to just two types. Without any loss of generality (wlog), we can assume that two gates are of the type $\mathcal{K}_{AB}$ and the third one is of the type $\mathcal{K}_{BC}$, and the circuit is (note that the time goes from left to right)
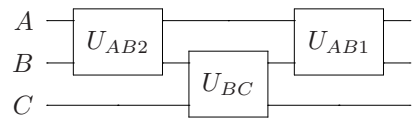
\[
\Qcircuit @C=1em @R=.7em {
\lstick{A} & \multigate{1}{U_{AB2}} & \qw & \multigate{1}{U_{AB1}} & \qw \\
\lstick{B} & \ghost{U_{AB2}}&  \multigate{1}{U_{BC}} & \ghost{U_{AB1}}&\qw\\
\lstick{C} & \qw & \ghost{U_{BC}} & \qw &\qw
}
\]
We only need to show there is no solution of the following equation
$$U_{AB1}U_{BC}U_{AB2}=V_{ABC},$$
where $U_{AB1}$ and $U_{AB2}$ are of type $\mathcal{K}_{AB}$, and $U_{BC}$ of type $\mathcal{K}_{BC}$. Then $U_{BC}$ must be a controlled gate with control on $C$ by noticing that $U_{BC}=U_{AB1}^{\dag}V_{ABC}U_{AB2}^{\dag}$, where $^{\dag}$ stands for the Hermitian conjugate. We can write
$U_{BC}=\op{0}{0}\otimes I_B+\op{0}{0}\otimes w_B$. A direct calculation leads us to the conclusion that $I_A \otimes w_B$ and $I-2\op{11}{11}$ share the same set of eigenvalues (counting multiplicity).
That is impossible.\\
Case 2: Three gates belong to different types. Wlog, we can assume the circuit is
$$U_{AB}U_{BC}U_{AC}=V_{ABC}.$$
We know that $U_{BC}U_{AC}$ is a controlled gate with control bit $C$. As just discussed, we can obtain that $U_{BC}$ is a controlled gate with control system $C$, so does $U_{AC}$. Consequently, we can assert that $I-2\op{11}{11}$ is local unitary by figuring directly out the form of control unitary. That is again impossible.

We can generalize this technique to show that the gate $V_{ABC}$ cannot implemented by any circuit consisting of four nonlocal two-qubit gates we do not count the number of one-qubit gates as they can be easily absorbed into relevant two-qubit gates. Again the symmetric property of $V_{ABC}$ enables us to consider only the following two cases:\\
Case 1: Four gates belong to only two types, say
$\mathcal{K}_{AB}$ and $\mathcal{K}_{BC}$. Due to the symmetry of $V_{ABC}$,
we only need to show the following circuit cannot be $V_{ABC}$,
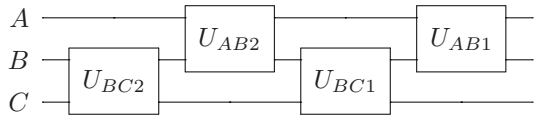
\[
\Qcircuit @C=1em @R=.7em {
\lstick{A} & \qw & \multigate{1}{U_{AB2}} & \qw &\multigate{1}{U_{AB1}}&\qw\\
\lstick{B} &  \multigate{1}{U_{BC2}}& \ghost{U_{AB2}} & \multigate{1}{U_{BC1}}& \ghost{U_{AB1}}&\qw\\
\lstick{C} & \ghost{U_{BC2}} & \qw &  \ghost{U_{BC1}} &\qw& \qw
}
\]
that is to show the following equation has no solution:
$$U_{AB1}U_{BC1}U_{AB2}U_{BC2}=V_{ABC}.$$
The proof detail of this case is given in appendix.

Case 2: Each of three types contains least one of the four two-qubit gates. Again due to the symmetry of $V_{ABC}$, we only need to deal with the following two subcases:\\
Case 2.1: The circuit is represented by $U_{AC}U_{AB1}U_{BC}U_{AB2}=V_{ABC}$. We can reduce this circuit to the circuit considered
in Case 1 by observing that $S_{AB}V_{ABC}S_{AB}=V_{ABC}$ and
$$
(S_{AB}U_{AC}S_{AB})(S_{AB}U_{AB1})U_{BC}(U_{AB2}S_{AB})=V_{ABC},
$$
where $S_{AB}$ is the swap gate on system $\mathcal{H}_A\otimes\mathcal{H}_B$ given by $S\ket{x_A}\ket{y_B}=\ket{y_A}\ket{x_B}$ for any two states $\ket{x}$ and $\ket{y}$.
Here we have employed the fact that $S_{AB}U_{AC}S_{AB}$ a two-qubit gate acting on $BC$, $S_{AB}U_{AB1}$ and
$U_{AB2}S_{AB}$ are two-qubit gates acting on $AB$.\\
Case 2.2: The circuit is represented by $U_{AB1}U_{BC}U_{AC}U_{AB2}=V_{ABC}.$ We know that $U_{BC}U_{AC}$ is a controlled gate with control system $C$. Directly, we can obtain $U_{BC}$ and $U_{AC}$ are controlled-gates with with control on $C$.
This leads us to the conclusion that $I-2\op{11}{11}$ shares eigenvalues counting multiplicity with a local unitary, which means
that the product of two eigenvalues of $I-2\op{11}{11}$ equals to the product of the other two. Impossible.

We have shown that four two-qubit gates are not sufficient for simulating the Toffoli gate, which further implies that any circuit consisting of less than five two-qubit gates has a positive distance to the Toffoli gate since the set of three-qubit gates that can be implemented by using up to four two-qubit gates form a compact set; in other words, the Toffoli gate cannot be well approximated by such circuits.

This above argument can also be used to show that the following two-qubit controlled phase gate (three-qubit quantum gates with two control systems and one target qubit) introduced by Deutsch \cite{DEU89} can not be implemented by four two-qubit gates:
\begin{eqnarray*}
V_{\theta}=I-(1-e^{i\theta})\op{111}{111}.
\end{eqnarray*}
where $0<\theta<2\pi$. Note that $V_{ABC}$ is the special case of $\theta=\pi$. Together with the result in \cite{SW95}, we conclude that five two-qubit gates are optimal for simulating the the two-qubit controlled phase gate.

In this paper, we study the problem of implementing multi-qubit gate using two-qubit unitaries. It is demonstrated that four two-qubit unitaries is not enough for constructing a three-qubit Toffoli gate, thus, five two-qubit gates is optimal. More precisely, our idea can be directly used to prove that in order to implement a three-qubit control phase gate, five two-qubit gates is also needed. We hope this work will be helpful for further determining minimal cost of implementing larger quantum logical gates, e.g. the multi-qubit controlled gate, and for studying optimization of quantum logical circuits, a crucial issue in the design and implementation of quantum computer hardware and architecture.

This work was partly supported by the Australian Research Council (Grant No:
DP110103473) and the Overseas Team Program of Academy of Mathematics and
Systems Science, Chinese Academy of Sciences.

\textbf{Technical appendix:} In this appendix, we show that there is no unitaries $U_{AB1},U_{AB2}$ and $U_{BC1},U_{BC2}$ such that
$$U_{AB1}U_{BC1}U_{AB2}U_{BC2}=V_{ABC}.$$ Notice that $U_{AB1}U_{BC1}U_{AB2}$ is a controlled gate on the bipartition $A-BC$ with control on $A$.
Moreover, the $A$'s part state of the output state
$U_{AB1}U_{BC1}U_{AB2}~\ket{i}_A\ket{\psi}_{BC}$
is still $\ket{i}_A$ for any input state $\ket{i}_A\ket{\psi}_{BC}$ with $i=0,1$.
Since $U_{AB2}$ maps some state $\ket{0}_A\ket{\xi}_B$ to product state, we can assume that
$U_{AB2}~\ket{0}_A\ket{0}_B=\ket{0}_A\ket{0}_B$ by absorbing one-qubit gates into $U_{BC1}$ and $U_{AB1}$. Then the state of A's part of $U_{AB1}U_{BC1}U_{AB2}~\ket{0}_A\ket{0}_B\ket{z}_C=U_{AB1}U_{BC1}~\ket{0}_A\ket{0}_B\ket{z}_C$ is still $\ket{0}_A$. We now need to consider three subcases according to different forms of the state $U_{BC1}\ket{0}_B\ket{y}_C$:\\
Case~~1.1: There is some $\ket{z_0}_C$ such that $U_{BC1}\ket{0}_B\ket{z_0}_C$ is entangled. Assume that there is $0<\lambda<1$ such that
$$U_{BC1}\ket{0}_B\ket{z_0}_C=\sqrt{\lambda}\ket{0}_B\ket{\alpha}_C+\sqrt{1-\lambda}\ket{1}_B\ket{\alpha^{\bot}}_C,$$
where we have absorbed a local unitary acting on $B$ into $U_{AB1}$.
Let $\ket{\Phi}=U_{AB1}\ket{00}$ and $\ket{\Psi}=U_{AB1}\ket{01}$, we know that
\begin{eqnarray*}
\ket{\chi}_{ABC}&=&U_{AB}U_{BC}\ket{0}_A\ket{0}_B\ket{z_0}_C \\
&=&\sqrt{\lambda}\ket{\Phi}_{AB}\ket{\alpha}_C+\sqrt{1-\lambda}\ket{\Psi}_{AB}\ket{\alpha^{\bot}}_C,
\end{eqnarray*}
we can readily obtain
\begin{eqnarray*}
\chi_A=\op{0}{0}=\lambda\Phi_A+(1-\lambda)\Psi_A \Rightarrow\Phi_A=\Psi_A=\op{0}{0}.
\end{eqnarray*}
Therefore, $U_{AB1}$ is a controlled gate with control system $A$, then one know that $U_{AB2}=U_{BC1}^{\dag}U_{AB1}^{+}V_{ABC}U_{BC2}^{\dag}$ is a controlled gate with control $A$. Assume that $U_{AB1}=\op{0}{0}\otimes I_B+\op{1}{1}\otimes u_B$ and
$U_{AB2}=\op{0}{0}\otimes I_B+\op{1}{1}\otimes v_B$. We conclude that
$U_{BC1}v_BU_{BC1}^{\dag}=u_B^{\dag}\otimes\op{0}{0}+u_B^{\dag}Z_B\otimes \op{1}{1},$ where $Z$ is the Pauli matrix given by $Z\ket{0}=\ket{0}$ and $Z\ket{1}=-\ket{1}$.

The set of eigenvalues of $U_{BC1}v_BU_{BC1}^{\dag}$ counting multiplicity is $\{e^{i\theta_1},e^{i\theta_1},e^{i\theta_2},e^{i\theta_2}\}$,
which is also the eigenvalues counting multiplicity of the right hand side of the above equality. Note that $u_B^{\dag}$ should not equal to identity up to a global phase. Then
$u_B^{\dag}$ and $u_B^{\dag}Z_B$ have the same set of eigenvalues. Thus their determinants are equal,
say \[\det(u_B^{\dag})=\det(u_B^{\dag}Z_B)=\det(u_B^{\dag})\det(Z_B)=-\det(u_B^{\dag}),\]
and $\det(u_B^{\dag})=0$. This contradicts the fact that $u_B^{\dag}$ is unitary.

Thus $U_{BC1}\ket{0}_B\ket{z}_C$ is always product for any $\ket{z}_C$. This leads us to consider the following two subcases.

Case~~1.2: There is a $\ket{\gamma}_C$ and a local unitary $w_B$ on system $B$
such that $U_{BC1}\ket{0}_B\ket{z}_C=w_B\ket{z}_B\ket{\gamma}_C$.
Then $U_{AB1}$ maps $\{\ket{0}_A\}\otimes \mathcal{H}_B$ to itself, hence $U_{AB1}$ is a controlled gate with control system $A$. Similarly $U_{AB2}$ is also a controlled gate with the same control bit. The rest proof is the same as Case 1.1.

Case~~1.3: There is a state on system $B$, wlog, says $\ket{0}_B$, and a local unitary $w_C$ on system $C$
such that $U_{BC1}\ket{0}_B\ket{z}_C=\ket{0}_B w_C\ket{z}_C$.
Then $U_{BC1}$ is a controlled gate with control system $B$. By moving this $w_C$ into $U_{BC2}$, we can assume that
$U_{BC1}=\op{0}{0}\otimes I_C+\op{1}{1}\otimes u_C.$
Note that for any $\ket{z}_C$, part $C$'s state of the output state
$\ket{\chi}_{ABC}=U_{AB1}U_{BC1}U_{AB2}~\ket{0}_A\ket{0}_{B}\ket{z}_C=U_{AB1}~\ket{0}_A\ket{0}_{B}\ket{z}_C$
is still $\ket{z}_C$. Recall that $\ket{\chi}_{ABC}=V_{ABC}\ket{0}_A (U_{BC2}^{\dag}\ket{0}_{B}\ket{z}_C)=\ket{0}_A (U_{BC2}^{\dag}\ket{0}_{B}\ket{z}_C)$.
Thus part $C$'s state of $U_{BC2}^{\dag}\ket{0}_{B}\ket{z}_C$ is $\ket{z}_C$ for all $\ket{z}_C\in\mathcal{H}_C$, which means that there is $\ket{\beta}_B$ such that $U_{BC2}\ket{\beta}_B\ket{z}_C=\ket{0}_B\ket{z}_C.$
Therefore, one can find a unitary $v_C$ such that $U_{BC2}=\op{0}{\beta}\otimes I_C+\op{1}{\beta^{\bot}}\otimes v_C.$
In order to simplify the structure of the two-qubit gates, we observe that
$U_{BC2}^{\dag}U_{AB2}^{\dag}U_{BC1}^{\dag}U_{AB1}^{\dag}=V_{ABC}$, $i.e.$, hence also provides a simulation of $V_{ABC}$.
Now we consider the state
$$U_{BC2}^{\dag}U_{AB2}^{\dag}U_{BC1}^{\dag}\ket{0}_C\ket{0}_B\ket{x}_A=U_{BC2}^{\dag}U_{AB2}^{\dag}\ket{0}_C\ket{0}_B\ket{x}_A$$ for any $\ket{x}_A$.
The argument of cases 1.1 and 1.2 excludes the following possibilities:
(i) there is some $\ket{x}_A$ such that $U_{AB2}^{\dag}\ket{0}_B\ket{x}_A$ is entangled, or (ii)
there is a $\ket{\delta}_A$ and a local unitary $w_B$ on system $B$ such that $U_{AB2}^{\dag}\ket{0}_B\ket{x}_A=w_B\ket{x}_B\ket{\delta}_A$. So the only possibility is that there is a state $\ket{\phi}_B$ on system $B$, and a local unitary $w_A$
on system $A$ such that $U_{AB2}^{\dag}\ket{0}_B\ket{x}_A=\ket{\phi}_B w_A\ket{x}_A$.
According to $U_{AB2}~\ket{0}_A\ket{0}_B=\ket{0}_A\ket{0}_B$, we can choose $\ket{\phi}=\ket{0}$.
Thus $U_{AB2}$ is a controlled gate with control system $B$, $i.e.$,
 $U_{AB2}=\op{0}{0}\otimes w_A+\op{1}{1}\otimes v_A.$
By studying part $C$'s state of $U_{AB1}U_{BC1}U_{AB2}U_{BC2}~\ket{0}_A\ket{0}_{B}\ket{z}_C=\ket{0}_A\ket{0}_{B}\ket{z}_C$,
we see that $\ket{\beta}_B$ defined in $U_{BC2}$ equals to $\ket{0}_B$ or $\ket{1}_B$,
up to some global phase. Otherwise, assume that $\ket{0}_B=a\ket{\beta}_B+b\ket{\beta^{\perp}}_B$ for $ab\neq 0$.
Then the state of part $C$ becomes a mixed state for general input $\ket{0}_A\ket{0}_{B}\ket{z}_C$ since $u_C$ is not identity up to some global phase and $U_{BC1}$ is nonlocal.
For the case $\ket{\beta}_B=\ket{0}_B$,
we know that all the four two-qubit gates are controlled gate with control system $B$, which implies that
$I-2\op{11}{11}$ is a local unitary, a contradiction.
For the case $\ket{\beta}_B=\ket{1}_B$, let $X_B$ be the NOT (flip) gate such that $X\ket{0}=\ket{1}$ and $X\ket{1}=\ket{0}$, then one can verify that
\[(U_{AB1}X_B)(X_BU_{BC1}X_B)(X_BU_{AB2}X_B)(X_BU_{BC2})=V_{ABC}.\]
Then $U_{AB1}X_B,X_BU_{BC1}X_B,X_BU_{AB2}X_B$ and $X_BU_{BC2}$
are all controlled gate with control system $B$.
This also leads us to the impossible conclusion that $I-2\op{11}{11}$ is local.

\begin{thebibliography}{99}

\bibitem{Sho97} Shor, P. SIAM J. Comp. \textbf{26}, 1484 (1997).

\bibitem{Fey82} Feynman, R. P. Inter. J. Theo. Phys. \textbf{21}, 467 (1982).

\bibitem{Hal02} Hallgren, S. Journal of the ACM, Volume 54 Issue 1, March 2007.

\bibitem{FKW02} Freedman, M. H., Kitaev, A. and Wang, Z. Comm. Math. Phys.  \textbf{227}, 587 (2002).

    \bibitem{NC00} Nielsen, M. A. and Chuang, I. L. Quantum Computation and
Quantum Information, First Edition (2000).

\bibitem{BDDG+2000} Bremner, M. J., Dawson, C. M., Dodd, J. L., Gilchrist, A., Harrow, A. W., Mortimer, D., Nielsen, M. A. and Osborne, T. J. Phys. Rev. Lett. \textbf{89}, 247902 (2002).

\bibitem{DCG+09} DiCarlo, L., Chow, J. M., Gambetta, J. M., Bishop, L. S.,
Johnson, B. R., Schuster, D. I., Majer, J., Blais, A., Frunzio, L.,
Girvin, S. M. and Schoelkopf, R. J. Nature \textbf{460}, 240 (2009).


\bibitem{MKH+09} Monz, T., Kim, K., Hansel, W., Riebe, M., Villar, A. S., Schindler, P., Chwalla, M., Hennrich, M. and Blatt, R. Phys. Rev. Lett. \textbf{102}, 040501 (2009).

\bibitem{LBA+09} Lanyon, B. P., Barbieri, M., Almeida, M. P., Jennewein, T., Ralph, T. C., Resch, K. J., Pryde, G. J., O'Brien, J. L., Gilchrist, A.  and White, A. G. Nature Phys. \textbf{5}, 134-140 (2009).

\bibitem{FSB+12} Fedorov, A., Steffen, L., Baur, M., da Silva, M. P. and Wallraff, A. Nature \textbf{481}, 170 (2012).


\bibitem{CMP+98} Cory, D. G., Mass, W., Price, M., Knill, E., Laflamme, R., Zurek, W. H., Havel, T. F. and Somaroo, S. S. Phys. Rev. Lett. \textbf{81},
2152-2155 (1998).

\bibitem{TOF80} Toffoli, T. in Automata, Languages and Programming, 7th Colloq. (eds de Bakker, J. W. van and Leeuwen, J.) 632-644 (Springer, 1980).

\bibitem{SHI03} Shi, Y. Quantum Inf. Comput. \textbf{3}, 84-92 (2003).

\bibitem{KLMN01} Knill, E., Laflamme, R., Martinez, R. and Negrevergne, C. Phys. Rev. Lett. \textbf{86}, 5811-5814
(2001).

\bibitem{CLS+04} Chiaverini, J., Leibfried, D., Schaetz, T., Barrett, M. D., Blakestad, R. B., Britton, J., Itano, W. M., Jost, J. D., Knill, E., Langer, C., Ozeri, R. and Wineland, D. J. Nature \textbf{432}, 602-605
(2004).

\bibitem{PJF05} Pittman, T. B., Jacobs, B. C. and Franson, J. D. Phys. Rev. A \textbf{71}, 052332 (2005).

\bibitem{AOK09}  Aoki, T., Takahashi, G., Kajiya, T., Yoshikawa, J., Braunstein, S., Loock, P. and Furusawa, A. Nature Phys. \textbf{5}, 541-546 (2009).
\bibitem{DEN01} Dennis, E.  Phys. Rev. A \textbf{63}, 052314 (2001).

\bibitem{MWY+11} Mariantoni, M., Wang, H., Yamamoto, T., Neeley, M., Bialczak, R. C.,
Chen, Y., Lenander, M., Lucero, E., O¡¯Connell, A. D., Sank, D., Weides, M., Wenner, J., Yin, Y., Zhao, J., Korotkov, A. N., Cleland, A. N. and Martinis, J. M. Science \textbf{334}, 61 (2011).



\bibitem{RDN+12} Reed, M. D., DiCarlo, L., Nigg, S. E., Sun, L., Frunzio, L., Girvin, S. M. and Schoelkopf, R. J. Nature \textbf{482}, 382-385 (2012).



\bibitem{DIV98} DiVincenzo, D.P. Proc. R. Soc. Lond. A, \textbf{454}:261-276,
(1998).

\bibitem{DS94} DiVincenzo, D.P. and Smolin, J. A.
Proc. of the Workshop on Physics and Computation (1994).

\bibitem{SW95} Sleator, T. and Weinfurter, H.  Phys. Rev. Lett. \textbf{74}, 4087 (1995).

\bibitem{BBCD+95} Barenco, A., Bennett, C. H., Cleve, R., DiVincenzo, D.P., Margolus, N., Shor, P., Sleator, T., Smolin, J. A., and Weinfurter, H. Phys. Rev. A \textbf{52}, 3457 (1995).

\bibitem{RRG07} Ralph, T. C., Resch, K. J., and Gilchrist, A. Phys. Rev. A 75, 022313 (2007).

\bibitem{MAR94} Margolus, N. Unpublished manuscript (circa 1994).

\bibitem{SK04} Song, G. and Klappenecker, A.  Quantum Inf. Comp. \textbf{4}, 361-372 (2004).

\bibitem{SM09} Shende, V. V. and Markov, I. L. Quantum Inf. Comp. \textbf{9}, 461 (2009).

\bibitem{DEU89} Deutsch, D. Proc. Roy. Soc. Lond. A. 425 (1989).

\end{thebibliography}
\end{document}